\documentclass[twocolumn]{webofc}

\usepackage[varg]{txfonts}   
\newcommand\numberthis{\addtocounter{equation}{1}\tag{\theequation}}
\begin{document}
\title{Variance minimisation on a quantum computer for nuclear structure}
%
%

\author{\firstname{Isaac} \lastname{Hobday}\inst{1}\fnsep\thanks{\email{i.hobday@surrey.ac.uk}} \and
        \firstname{Paul} \lastname{Stevenson}\inst{1}\fnsep\thanks{\email{p.stevenson@surrey.ac.uk}} \and
        \firstname{James} \lastname{Benstead}\inst{2}
}

\institute{Department of Physics, University of Surrey, Guildford, GU2 7XH, UK 
\and
           AWE, Aldermaston, Reading, RG7 4PR, UK 
          }

\abstract{%
  Quantum computing opens up new possibilities for the simulation of many-body nuclear
systems.
As the number of particles in a many-body system increases, the size of the space if the associated
Hamiltonian increases exponentially. This presents a challenge when performing calculations
on large systems when using classical computing methods. By using a quantum computer,
one may be able to overcome this difficulty thanks to the exponential way the Hilbert space
of a quantum computer grows with the number of quantum bits (qubits).
Our aim is to develop quantum computing algorithms which can reproduce and predict
nuclear structure such as level schemes and level densities. As a sample Hamiltonian, we
use the Lipkin-Meshkov-Glick model.
We use an efficient encoding of the Hamiltonian onto many-qubit systems, and have
developed an algorithm allowing the full excitation spectrum of a nucleus to be determined
with a variational algorithm capable of implementation on today’s quantum computers with
a limited number of qubits. Our algorithm uses the variance of the Hamiltonian, $\left<H^2\right> - \left<H\right>^2$,
as a cost function for the widely-used variational quantum eigensolver (VQE).
In this work we present a variance based method of finding the excited state spectrum of a small nuclear system using a quantum computer, using a reduced-qubit encoding method.
}
\maketitle
\section{Introduction}
\label{sec-intro}

Quantum computers can theoretically provide significant advantages for certain computational tasks. In one kind of quantum computer, operations are performed on quantum bits (qubits) to perform calculations. Simulation of Fermionic systems is acheivable due to the spin-$\frac{1}{2}$ nature of qubits, and was one of the first proposed uses for quantum computers \cite{feynman_simulating_1982}. This is possible using  algorithms such as quantum phase estimation \cite{Guzik2005}, or quantum time evolution \cite{McArdle2019}, but these algorithms are too computationally expensive to be performed on current quantum computers. Variational algorithms allow some calculations of small nuclear systems to be performed on current hardware.

The uses of the Variational Quantum Eigensolver (VQE) are well documented \cite{Peruzzo2014,McClean_2016,tilly2022variational}. The VQE can be used to solve for the ground state of nuclear Hamiltonians, achieved by using the classical computer to minimize the energy expectation value of the Hamiltonian. If the optimizer is instead set to minimize the variance of the Hamiltonian,

\begin{equation}
    \sigma^2 = \left<H^2\right> - \left<H\right>^2,
\end{equation}

then the VQE can be used to find the excited state spectra of a nuclear model, since at points where this variance is equal to zero an eignestate is found.

\section{Lipkin-Meshkov-Glick Model}
\label{sec-LMG}
The Lipkin-Meshkov-Glick (LMG) model \cite{LIPKIN1965188} is a two-level nuclear shell model, where the two levels are separated by some energy, $\epsilon$. The LMG Hamiltonian in the second-quantization form is 

\begin{align*}
    H = &\frac{1}{2}\epsilon \sum_{p\sigma}\sigma a_{p\sigma}^\dagger a_{p\sigma} + \frac{1}{2}V\sum_{pp'\sigma} a_{p,\sigma}^\dagger a_{p',\sigma}^\dagger a_{p',-\sigma}a_{p,-\sigma} \\&+ \frac{1}{2}W \sum_{pp'\sigma}a_{p,\sigma}^\dagger a_{p',-\sigma}^\dagger a_{p',\sigma} a_{p,-\sigma} \label{eq:LipkinOri}. \numberthis{}
\end{align*}

Here, $p$ and $\sigma$ correspond to the particle number and the energy level respectively. $V$ is the interaction strength for a pair of particles scattering from one level to the other, and $W$ is the interaction strength of one particle scattering from the lower level to the upper level and one particle scattering from the upper to the lower energy level.

The LMG model Hamiltonian can be represented in the quasispin basis, which reduces the size of the Hamiltonian matrix from order $2^N$ to order $N+1$. The LMG Hamiltonian in the quasispin representation is given as 

\begin{equation}
    H = \epsilon J_z + \frac{1}{2}V(J_+^2 + J_-^2) + \frac{1}{2}W(J_+J_- + J_-J_+)\label{eq:LipRed},
\end{equation}

where $J_z$ and $J_\pm$ are given by 

\begin{equation}
    J_z = \frac{1}{2}\sum_{p\sigma} \sigma a_{p\sigma}^\dagger a_{p\sigma} \label{eq:LipJz}
\end{equation}

and 
\begin{equation}
    J_\pm = \sum_p a_{p\pm1}^\dagger a_{p\mp 1} \label{eq:LipJ+}
\end{equation}

respectively.

For this work, we consider the $N=3$ case, where $N$ represents the number of particles in the model. This $N=3$ Hamiltonian corresponds to a $4\times4$ matrix, which is the size of matrix that can be fully utilised by two qubits. The Hamiltonian matrix in the quasispin basis is given as

\begin{equation}
    H_{N=3} = \begin{bmatrix}
    -1.5 & 0 & -0.866 & 0\\
    0 & -0.5 & 0 & -0.866 \\
    -0.866 & 0 & 0.5 & 0 \\
    0 & -0.866 & 0 & 1.5 \\
    \end{bmatrix}. \label{LMG_matrix}
\end{equation}

This matrix uses the values $\frac{V}{\epsilon} = 0.5$ and $\frac{W}{\epsilon}=0$, with all energies in units of $\epsilon$. LMG Hamiltonians created in the quasispin basis are block-diagonal. However, the matrix presented in Equation \ref{LMG_matrix} will be treated as a $4\times4$ matrix for this work, rather than two separate $2\times2$ matrices, to avoid the problem becoming too trivial on a quantum computer since a $2\times2$ matrix requires only a single qubit to represent the matrix space. To perform calculations using this Hamiltonian on a quantum computer one must represent the matrix in the form of Pauli spin matrices,

\begin{equation}
    H_{N=3} = -1.0 Z_0I_1 - 0.5 I_0Z_1 - 0.866 X_0I_1. \label{eq-LMG_pauli}
\end{equation}

Here the subscript after the Pauli spin matrix represents the qubit which that matrix is applied to. $Z_0I_1$ represents the tensor product of the $Z$ and $I$ Pauli matrices, $Z\otimes I$. This decomposition is created using a reduced encoding method \cite{Hobday,pesce_h2zixy_2021}, which minimizes the number of qubits required by the quantum computer. This method is, therefore, more qubit-efficient than the typically used Jordan-Wigner encoding method \cite{Jordan1928}.

Due to its exactly-solvable nature, the LMG model is very well suited as a test platform for quantum computing algorithms. Calculation of the ground state has been performed in the quasispin basis of the LMG model using IBM quantum computers \cite{cervia_lipkin_2021}.

\section{Simulation and Measurements}

\begin{figure}[h]
\centering
\includegraphics[width=7.5cm,clip]{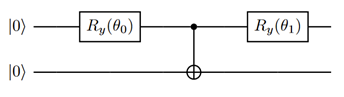}
\caption{Two-parameter ansatz circuit.}
\label{fig-circ}       
\end{figure}

The variational ansatz shown in Figure \ref{fig-circ} is used in the simulation and measurement of the LMG Hamiltonian given in Equation \ref{eq-LMG_pauli}. This is a low depth, two parameter circuit performed using two qubits, which is able to cover enough of the Hilbert space to find all eigenstates of the nuclear model. There are two \texttt{$R_y$} gates within this quantum circuit, which each perform a rotation of a given angle about the $y$ axis for a single qubit. These \texttt{$R_y$} gates are separated by a controlled-NOT (CNOT) gate, which creates entanglement between the two qubits in the circuit.
 
\begin{figure}[h]
\centering
\includegraphics[width=8.5cm,clip]{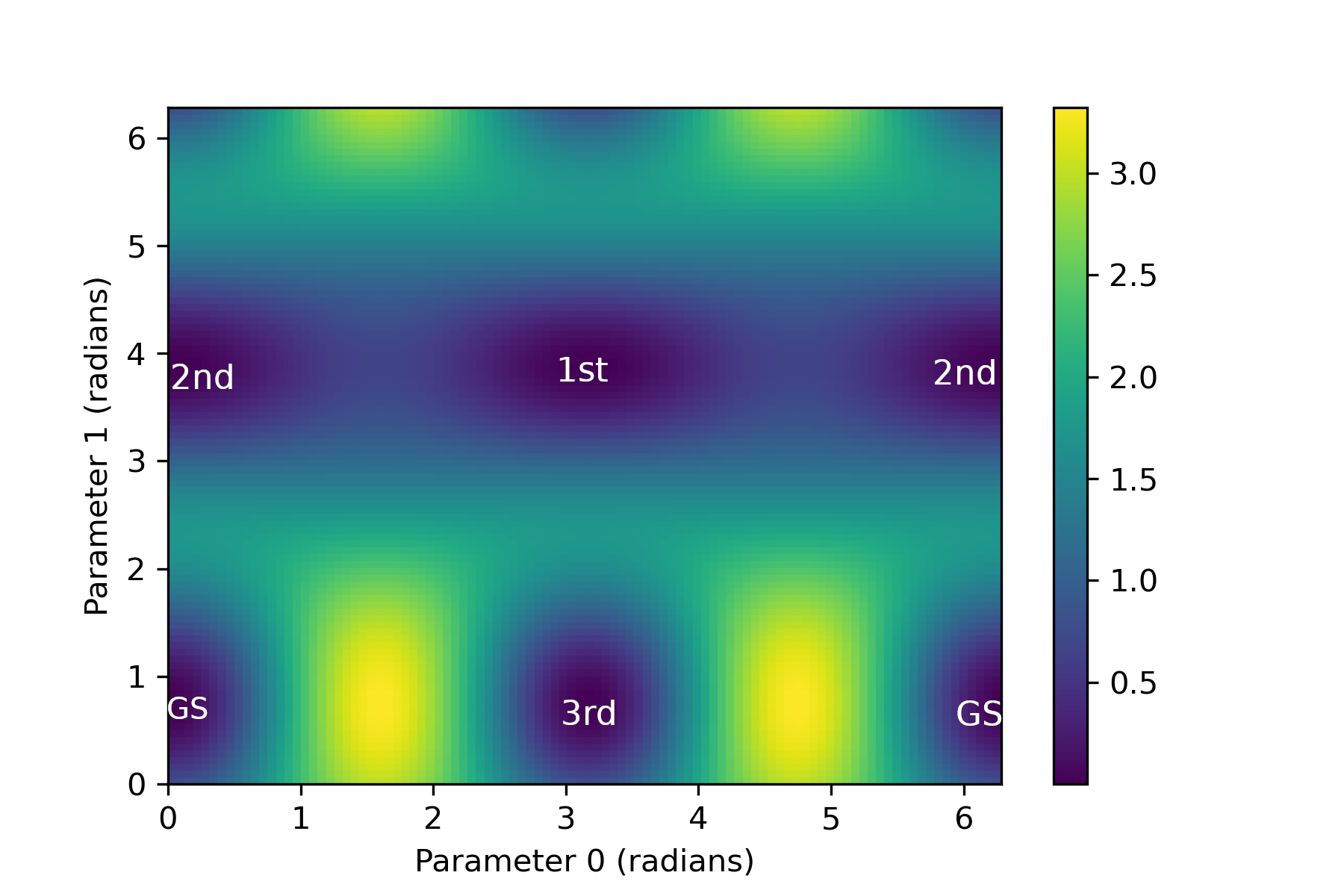}
\caption{Variance landscape for $N=3$ LMG model.}
\label{fig-variance}       
\end{figure}

Using a statevector simulator we can perform a sweep of the complete parameter space of the system. This creates the plot in Figure \ref{fig-variance}. in this plot it is clear that there are four minima present, corresponding to the four eigenstates of the Hamiltonian. The minima have been labelled with their respective states. Due to the cyclical nature of the variational parameters the function repeats when the value of the parameters exceeds $2\pi$.

Simulation provides a good idea of how a quantum computer will perform, however for significantly larger systems where greater numbers of qubits are required simulation will not be possible due to the exponentially increasing computational resource requirements. At the point where the size of the problem scales beyond the computational ability of classical computers, i.e. becomes intractable, quantum computation may provide the solution, due to the exponentially scaling nature of the Hilbert space when additional qubits are introduced.

We perform calculations using \textit{ibmq\_manila}, a 5-qubit quantum computer accessible via the cloud. As the wavefunction of the quantum circuit collapses into a single state once measured, it is important to measure the circuit a large number of times to determine the true state of the qubits prior to measurement. Therefore, calculations are performed using 20,000 measurements, also known as shots, of each quantum circuit. 

\section{Results}
\label{sec-Res}

The results of the variance VQE are shown in Table \ref{tab-1} where they are compared to the exact values of the eigenstates. These results have readout-bias correction applied, mitigating some affects of measurement error within the quantum computer \cite{Kandala2017}. 

\begin{table}[h]
\centering
\caption{Variance QC Results.}
\label{tab-1}       
\begin{tabular}{lll}
\textbf{Eigenstate} & \textbf{Exact Value (MeV)} & \textbf{QC Result (MeV)}  \\\hline
Ground & -1.823 & -1.798 \\ \hline
1st & -0.823 & -0.818\\\hline
2nd & 0.823 & 0.834\\\hline
3rd & 1.823 & 1.819\\\hline
\end{tabular}
\end{table}

Error calculations are still to be confirmed.
These results, determined by the quantum computer, show good accuracy when compared to the known exact values of the eigenstates for the $N=3$ LMG Hamiltonian.

\section{Conclusion}
\label{sec-Con}
We have presented results of VQE calculations in which we minimize the variance of the Hamiltonian. This method is shown to provide good results for the excited state spectrum of a small LMG Hamiltonian, both via quantum-mechanical methods and when performed on a quantum computer.

Further calculations using this method will be applied to larger LMG model Hamiltonians to determine how the algorithm will scale with increasing size of Hamiltonian. The method may then be applied to more complex shell-model calculations.

\section{Acknowledgements}
\label{sec-Ack}

This work was funded by AWE. We acknowledge the use of IBM Quantum services for this work. The views expressed are those of the authors, and do not reflect the official policy or position of IBM or the IBM Quantum team. In this paper we used \textit{ibmq\_manilla}, which is one of the IBM Quantum Falcon Processors. UK Ministry of Defence \copyright  Crown owned copyright 2022/AWE.\newline

%
\bibliography{bib.bib}
%
%
%
%


\end{document}